%% file: gamma_new.tex
\documentclass[epj]{svjour}
\usepackage{graphics,color}
\usepackage[T1]{fontenc}
\usepackage{graphicx}
\usepackage{amsmath,amssymb,bm}%,amsthm,mathtools}
\usepackage[draft]{fixme}
\usepackage{booktabs}
\usepackage{amsmath}
\usepackage{colortbl}
\usepackage{relsize}
\usepackage{sidecap}

\usepackage{graphicx}
\usepackage{caption}
\usepackage{textcomp}
\usepackage{gensymb}

\usepackage{lipsum}
\newcommand{\lambdabar}{{\mkern0.75mu\mathchar '26\mkern -9.75mu\lambda}}
\begin{document}

\title{Unbound states in $^{12}$C populated by $\gamma$-decay of the $(J^{\pi},T) = (2^+,1)$ 16.11~MeV state}
\author{K. L. Laursen\inst{1} \and H. O. U. Fynbo\inst{1}\thanks{e-mail: fynbo@phys.au.dk}  \and O. S. Kirsebom\inst{1} \and K. S. Madsb\o ll\inst{1} \and K. Riisager\inst{1}
} % Do not remove
\offprints{}          % Insert a name or remove this line
\institute{Department of Physics and Astronomy, Aarhus University, DK-8000 Aarhus C, Denmark}
\date{Received: date / Revised version: date}
\abstract{ The reaction $^{11}\textrm{B}+p$ has been used to populate the $(J^\pi,T) = (2^+,1)$ state at an excitation energy of 16.11\,MeV in $^{12}$C. $\gamma$-decay to unbound states in $^{12}$C are identified from analysis of the decay of the populated daughter states. Due to a new technique, $\gamma$-decay to the 10.8\,MeV 1$^-$ state is observed for the first time, and transitions to the 9.64\,MeV (3$^-$) and 12.71~MeV (1$^+$) are confirmed. Unresolved transitions to natural parity strength at 10\,MeV and 11.5-13\,MeV are also observed. For all transitions partial widths are deduced. 
 % end of PACS codes
} %end of abstract
%
%\authorrunning
\titlerunning{Unbound states in $^{12}$C populated by $\gamma$-decay of the $(J^{\pi},T) = (2^+,1)$ 16.11~MeV state}
\maketitle
\input{introduction}
\input{methodology}
\input{2p-review}

\input{experiment}
\input{analysis}
\input{discussion}

\section*{Acknowledgements} 
This work has been supported by the European Research Council under ERC starting grant LOBENA, No. 307447. OSK acknowledges support from the Villum Foundation.\\


\end{document}

%% file: introduction.tex
\section{Introduction}
\label{sec:introduction}
The spectrum of $^{12}$C has been the subject of countless studies throughout the history of nuclear physics. The nucleus has only two
bound states with the continuum opening at the 3$\alpha$ threshold slightly above 7\,MeV. Much work has focused on understanding the
continuum between the 3$\alpha$ threshold and the proton threshold near 16\,MeV. The first resonance is already at 7.65\,MeV (0$^+$), the
so called ``Hoyle state'', which is dominating the rate of the triple-$\alpha$ reaction in stars. Above that there are well
identified 3$^-$, 1$^-$, 2$^-$, 1$^+$, and 4$^+$ resonances below the proton threshold~\cite{ajzenberg90}.

In the past decade experimental evidence has mounted for additional resonances in this energy range. From a study of the
$^{12}$C($^{12}$C,3$\alpha$)$^{12}$C reaction Freer {\it et al.} found  evidence for broad 1$^{-}$ and 3$^{-}$ resonances at 11.8 and 12.5 MeV, while the resonance at 13.35\,MeV listed as 2$^-$ in~\cite{ajzenberg90} was reassigned as 4$^-$~\cite{freer07}; the latter was confirmed by Kirsebom {\it et al.}~\cite{kirsebom10}. Evidence for a 2$^+$ state near 10\,MeV has come from inelastic
$\alpha$- and proton- scattering \cite{freer09,zimmerman11,itoh11,freer12}, transfer reactions \cite{smit12} and $\gamma$ dissociation of $^{12}$C
\cite{zimmerman13,zimmerman13b}. R-matrix analysis of data from the $\beta$-decay of $^{12}$B and $^{12}$N suggests overlapping $0^+$ and 
$2^+$ structures at a somewhat higher energy of 11.1\,MeV, but provided no support for 2$^+$ strength at 10\,MeV~\cite{hyldegaard10}. Finally, 
evidence for a $4^+$ resonance at 13.3\,MeV~\cite{freer11}, and for a broad unnatural parity state with J$\geq$4 at 12.4\,MeV~\cite{kirsebom13} 
have been reported. 

As recently reviewed~\cite{freer14}, there is also theoretical support for additional states in the spectrum of $^{12}$C between the 3$\alpha$- and proton-thresholds. Many models either predict or assume cluster structure of $^{12}$C. An example is the 
algebraic cluster model of Bijker and Iachello~\cite{bijker00}, which recently received experimental verification~\cite{lambarri14}. This model
predicts several states between 7 and 16\,MeV including degenerate pairs of parity doublet states.

The main challenge with clarifying this situation is that many of the new resonances suggested either by data or theory have widths in excess of
an MeV and are overlapping. There is therefore a need for new experimental methods which selectively can populate specific J$^{\pi}$ states such that 
the individual contributions from overlapping states can be resolved.

Kirsebom {\it et al.} has developed a method for identifying $\gamma$-decay to unbound states in $^{12}$C~\cite{kirsebom09}, see also~\cite{alcorta09}. The method has been applied successfully in a proof-of-principle study using the 1$^+$ states in $^{12}$C at 12.71\,MeV and 15.11\,MeV as parent states for the $\gamma$-decay~\cite{kirsebom09}. This method resembles the method of $\beta$-delayed particle emission both in its selectivity and in the nature of the measured spectra. In particular this is the case for the spectrum from the 15.11\,MeV state because it is a member of the same isospin multiplet as the ground states of $^{12}$B and $^{12}$N, and the effective M1 operator responsible for the $\gamma$-decay is closely related to the effective Gamow-Teller operator responsible for the $\beta$-decay. 

Here we report on a measurement of the $\gamma$-decay of the ($2^+$, T=1) state at 16.11\,MeV in $^{12}$C to unbound states. E1 and M1 transitions
from this state can populate J=1, 2 and 3 states with negative and positive parity, respectively. This measurement may therefore cast some light on 
resonances with these spins and parities.

The paper is structured in the following way: section \ref{sec:methodology} describes the method of $\gamma$-decays to unbound states. Then,  section~\ref{sec:2p-review} reviews the previous work on the decay of the 16.11\,MeV state, and section \ref{sec:exp} explains the experimental setup
including geometry and energy calibrations. Section \ref{sec:analysis} presents the measured data, accounts for important parts of the applied analysis, and presents the experimental results, and finally in section~\ref{sec:discussion} the results of the measurement are discussed.

Other aspects of the experiment discussed here have been recently published~\cite{laursen16}.

%% file: methodology.tex
\section{Methodology}
\label{sec:methodology}

Detecting electromagnetic transitions to or from broad particle-unbound resonances with widths in excess of a few tens of keV is complicated due to the large spread in $\gamma$ energy. Also, because $\gamma$-decay widths are limited to a few eV or less, the $\gamma$-decay branching ratio for states in the continuum are typically less than 10$^{-3}$. Indirect detection of $\gamma$-decay is a method developed for these cases. The basic idea is to substitute direct $\gamma$-ray detection with the detection of the particle decay of the unbound state populated after the $\gamma$-decay. By knowing the energy of the parent state, the $\gamma$-decay is identified by the missing energy that equals the difference in energy between the parent and daughter levels. 
 
This method offers several advantages compared to conventional methods for $\gamma$ spectroscopy. Charged particle detectors (see Section \ref{sec:exp} for the detectors used here) have a simpler response function than modern $\gamma$-ray detectors, which makes it more feasible to measure the non-Gaussian profile of a broad resonance. Also, the intrinsic efficiency of $\gamma$-ray detectors limits their applicability for weak decay branches. If the natural width of the unbound state is several tens of keV the reduced resolution of charged particle detectors compared to high resolution $\gamma$-ray detectors is of little consequence.
 
The case of interest here are resonances in $^{12}$C, which decay by triple-$\alpha$ emission. This offers a challenge and additional advantages. The challenge is to design a detection system capable of detecting with high efficiency final states of three particles. This is today relatively simple to achieve with large area, segmented Silicon detectors, as will be described in section~\ref{sec:exp} for the experiment discussed here. The advantages include the fact that recording the decay of the daughter state in triple coincidence essentially is background free since many analysis conditions can be placed on such events in the off-line analysis. In addition kinematic analysis of the triple-$\alpha$ decay can split the events into decays occurring via the 0$^+$ ground state in $^8$Be, and decays occurring via higher energies in the two-$\alpha$ system. Since angular momentum and parity conservation forbids unnatural parity states to decay through the ground state of $^8$Be, this provides an additional spectroscopic tool. In the present case this method will help the assignment of the final states populated by either E1 or M1 transitions.

Partial $\gamma$ widths can be calculated by measuring the branching ratio $B_{\gamma,i}$ of the individual $\gamma$-decays and normalising to the to total width of the parent state $\Gamma_\textrm{tot}$,  
\begin{align}
\Gamma_{\gamma,i} = \Gamma_\textrm{tot} B_{\gamma,i}.
\label{gammagamma1}
\end{align}
As usual, the branching ratio can be determined as the ratio between the total number of decays of the parent state to the number of decays to a particular final state following the $\gamma$-decay. 
  
The method of indirect detection of $\gamma$-rays is further discussed in~\cite{kirsebom09,alcorta09,kirsebom14}.

%% file: 2p-review.tex
\section{Properties of the 16.11 MeV 2$^+$ resonance}
\label{sec:2p-review}
\begin{figure*}
\centering
\includegraphics[width=0.8\textwidth]{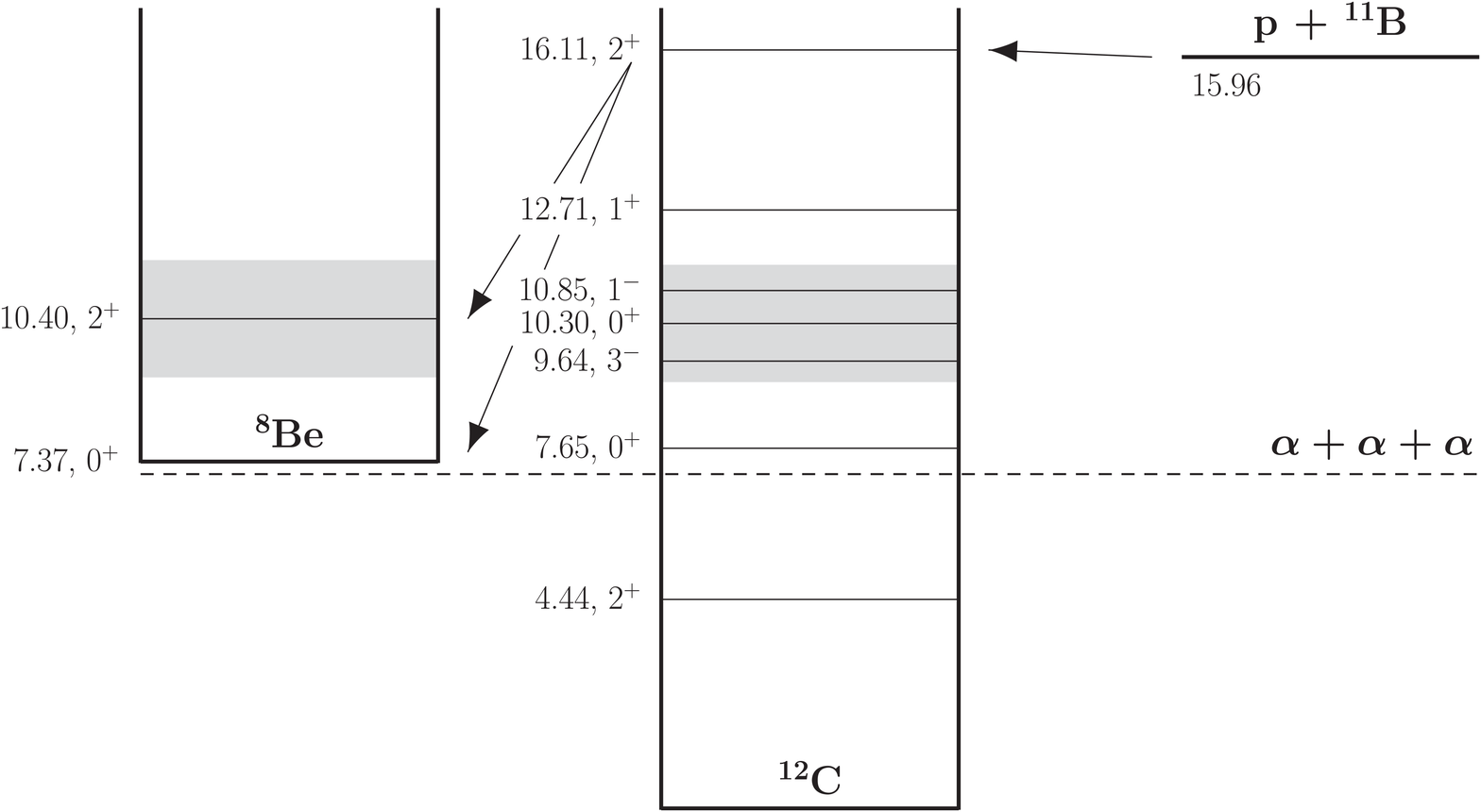}
\caption{\label{fig:scheme}Decay scheme for $^{12}$C the 16.11 MeV state. All energies are in MeV relative to the ground state of $^{12}$C. Above the triple-$\alpha$ threshold only states of interest are shown, all taken from the review of Ajzenberg-Selove and Kelley~\cite{ajzenberg90}. Note, the 10.3~MeV 0$^+$ state has later been reinterpreted~\cite{hyldegaard10,fynbo05,diget05}. The shading (grey) indicates regions with broad resonances ($\Gamma \,{\sim}1\,$MeV).}
\end{figure*}

\begin{table*}[h]
\centering
\caption{\label{tab:partialwidths-review} Partial widths of the 16.11 MeV resonance from the latest evaluation \cite{ajzenberg90}, and updated values as discussed in the text.}
\begin{tabular}{c|c|c|c} 
\toprule
Channel & $\Gamma_x$ (\cite{ajzenberg90}) & $\Gamma_x$ (updated) & Comments \\ 
\hline
$\Gamma_{tot}$   & 5.3(2) keV & 5.3(2) keV     & Combination of \cite{davidson79,becker87} \\
\hline
$\alpha_0$       & 290(45)   eV & 270(30) eV   & From total width and \\ 
$\alpha_1$       & 6.3(5)   keV & 5.0(2) eV    & $\frac{\Gamma_{\alpha0}}{\Gamma_{\alpha1}}$ from~\cite{laursen16} \\ 
\hline
p                & 21.7(1.8) eV & 37(7) eV     & From $\sigma_{p\gamma0}$ and $\sigma_{p\gamma1}$ and eq.~\ref{sigmax} \\ 
\hline
$\gamma$ (gs)    & 0.59(11)  eV & 0.35(4) eV   & From electron scattering \cite{friebel78} \\
\hline
$\gamma$ (4.44)  & 12.8(1.7) eV & 10.5(1.6) eV & Combination of $\Gamma_{tot}$~\cite{davidson79,becker87} and $\frac{\Gamma_{\gamma}(4.44)}{\Gamma_{\alpha}}$~\cite{cecil92}.\\
\hline
$\gamma$ (9.64)  & 0.31(6)   eV & 0.25(7) eV   & Combination of $\Gamma_{\gamma}(4.44)$ and \\
$\gamma$ (12.71) & 0.19(4)   eV & 0.15(3) eV   & $\frac{\Gamma_{\gamma}(9.64,12.71)}{\Gamma_{\gamma}(4.44)}$ \\
\bottomrule
\end{tabular}
\end{table*} 
In the present work the $p\,+\,^{11}\textrm{B}$ reaction is used to populate the $\left[J^\pi = 2^+, T = 1\right]$ state at 16.11\,MeV in $^{12}$C. The known decay branches of this state are proton emission, $\alpha$-emission, and $\gamma$ decays to the ground state, the $2^+$ state at 4.44\,MeV, the $3^-$ state at 9.64\,MeV, and to the $1^+$ state at 12.71\,MeV. The relevant data has been reviewed and evaluated by Ajzenberg-Selove and Kelley~\cite{ajzenberg90}. Figure~\ref{fig:scheme} shows the known decay branches.
 
The literature on the partial decay branches of the 16.11~MeV state is quite rich including absolute and relative measurements of the $\gamma$- and $\alpha$-emission branches. These values are often in mutual disagreement, and it is therefore challenging to combine the previous measurements into recommended values. However, it seems several of the partial widths should be updated compared to the latest evaluation~\cite{ajzenberg90}. This is partly caused by the appearance of new data. 

The $\alpha$-emission channels have been the subject of many studies. It is established that approximately 5\% of the $\alpha$-decays go to the ground state of $^8$Be (referred to as the $\alpha_0$-channel). The interpretation of the remaining 95\% of the $\alpha$-decays (referred to as the $\alpha_1$-channel) has been contested. The analysis of these channels from the same experiment as discussed in the present paper has been recently published including a new, precise measurement of the ratio of the partial $\alpha$-decay branches $\frac{\Gamma_{\alpha0}}{\Gamma_{\alpha1}}$=5.1(5)\%~\cite{laursen16}. 

In the narrow resonance limit the partial widths $\Gamma_x$ are related to partial cross sections $\sigma_{px}$ in the reaction $^{11}$B(p,x) by the relation
\begin{equation}
    \sigma_{px} = 4 \pi \lambdabar^2 \omega \Gamma_p \Gamma_x / \Gamma_{tot}^2.
\label{sigmax}
\end{equation}
In this expression, the total width $\Gamma_{tot}$ is measured by both Davidson {\it et al.} \cite{davidson79} and by Becker {\it et al.}~\cite{becker87} leading to a combined value of $\Gamma_{tot}$=5.3(2)~keV. The ground state $\gamma$-decay width $\Gamma_{\gamma}(gs)$=0.35(4)eV~\cite{friebel78} has been measured by electron scattering. Electron scattering constitutes a direct measurement of the ground state radiative width, and therefore ought to be more reliable than indirect approaches. We therefore adopt the value from electron scattering here, although it is almost half the value recommended by the evaluation~\cite{ajzenberg90}.

The new value for the ratio $\frac{\Gamma_{\alpha0}}{\Gamma_{\alpha1}}$ from~\cite{laursen16} together with the value for the total width leads to new values for the partial $\alpha$-decay branches $\Gamma_{\alpha0}$ and $\Gamma_{\alpha1}$ given in table~\ref{tab:partialwidths-review}. 

The cross sections $\sigma_{p\gamma0}$ and $\sigma_{p\gamma1}$ for producing the ground state and first excited state $\gamma$-rays have been consistently measured both by Huus {\it et al.}~\cite{huus53}, and by Cecil {\it et al.} \cite{cecil92} with combined values of $\sigma_{p\gamma0}$=5(1)~$\micro$b and $\sigma_{p\gamma1}$=140(20)~$\micro$b respectively. Combining these values and eq.~\ref{sigmax} gives $\Gamma_{p}$=37(7)~eV. This value is almost a factor 2 larger than that of~\cite{ajzenberg90}, and although less precise seems more reliable. Using this value and eq.~\ref{sigmax} leads to $\sigma_{p\alpha}$=75(15)~mb, which is consistent with the measurement of Becker {\it et al.}~\cite{becker87}.   

The partial width for the $\gamma$-decay to the 4.44\,MeV state can be obtained by combining the total width~\cite{davidson79,becker87} with the value of $\frac{\Gamma_{\gamma}(4.44)}{\Gamma_{\alpha}}$=2.0(3)$\times$10$^{-3}$~\cite{cecil92}. The partial $\gamma$-decay branches to the excited states at 9.64~MeV and 12.71~MeV were measured relative to the transition to the 4.44~MeV state by Adelberger {\it et al.}~\cite{adelberger77}. This can be used to derive the partial widths to these states given in table~\ref{tab:partialwidths-review}.

%% file: experiment.tex
\section{Experimental arrangement}
\label{sec:exp}
The experiment to be described was conducted during a period of 6 months. It consisted in producing the 16.11\,MeV state by proton bombardment of $^{11}$B, and detecting $\alpha$-particles from decay of the state with Silicon detectors. 

\begin{figure}
\includegraphics[width=0.8\columnwidth]{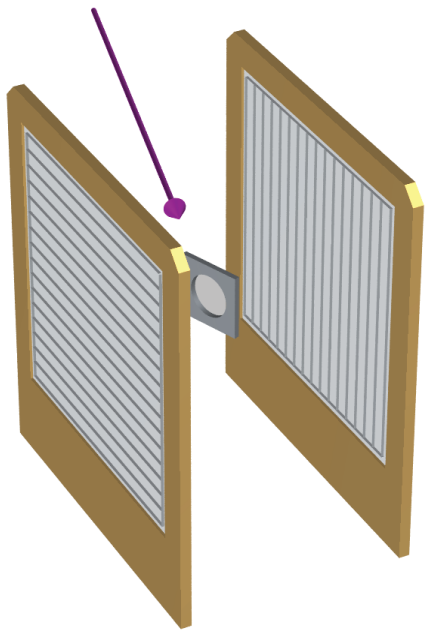}%Geo1_newFINAL}
\caption{Detector setup used in the experiment; DSSSD1 is to the right, DSSSD2 is to the left. Each detector has a size of 50$\times$50\,mm$^2$ and they are separated by approximately 50\,mm. The target is in the centre and the beam direction is illustrated by the arrow.}
\label{fig:det_setup}
\end{figure}

The proton beam from the 400\,keV Van de Graaff accelerator at Aarhus University was passed through a magnetic analyzer to bombard thin natural boron targets with thicknesses ranging between 10-15\,$\mu \textrm{g/cm}^2$ on 4\,$\mu \textrm{g/cm}^2$ carbon backing. With a set of horizontal and vertical slits the beam was reduced to a size of 2\,mm $\times$ 2\,mm and beam currents of typically 1\,nA. The detection system consisted of two double-sided silicon strip detectors (DSSSD1 and DSSSD2) of the W1 type with 16 $\times$ 16 strips and an active area of 5 cm $\times$ 5 cm. DSSSD1 had a thin deadlayer equivalent to 200\,nm Si, while the deadlayer of DSSSD2 was 700\,nm.  Both detectors were 60\,$\mu\textrm{m}$ thick; enough to fully stop all $\alpha$ particles from the $p\,+\,^{11}\textrm{B}$ reaction. For the largest part of the experiment the detectors were positioned as shown in Fig.~\ref{fig:det_setup} covering center-of-mass polar angles in the intervals $60\degree-150\degree$ (DSSSD 1) and $35\degree-120\degree$ (DSSSD 2). 

The electronics chain consisted of Mesytec MPR-32 charge-sensitive preamplifiers, and Mesytec STM16+ shaping amplifiers and discriminators. The shaped and amplified signals were digitized with CAEN 785 analogue-to-digital-converters (ADC), and the discriminator outputs were used to generate the trigger, and after a delay fed to a CAEN 1190 time-to-digital converter (TDC). The amplification gain was stable throughout the experiment. The energy resolution of the system for both detectors was 40\,keV (FWHM), and the time resolution was approximately 100\,ns. The trigger efficiencies were found to rise gently as a function of energy, increasing from 0\% to 100\% within an interval of 200-400\,keV. Trigger thresholds, defined as the energy at which the efficiency reaches 50\%, ranged from 100 to 300\,keV for DSSSD 1, and from 200 to 500\,keV for DSSSD 2. Low energy cutoffs in each ADC channel ranged from 10 to 100\,keV for DSSSD 1 and from 100 to 200\,keV for DSSSD 2. Further information on the experimental arrangement can be found in~\cite{laursen16,laursenPhd}.

For energy calibration we used the six most intense $\alpha$ lines from a $^{228}$Th source, which ranches from 5.4 to 8.8\,MeV. Throughout the experiment calibrations were made regularly. SRIM \cite{SRIM} range tables were used to correct for energy losses in the source itself and the in detector deadlayers. Corrections were also made for non-ionizing energy losses \cite{Lennard1986} in the active detector volume. By measuring with the source in rotated positions its thickness was determined to be equivalent to 100(4)\,nm of graphite. The detector deadlayer thicknesses were found by studying the variation of pulse height across a strip due to the change in effective deadlayer thickness. 

During the 6 months used for the experiment several smaller changes to the setup was made. The total acquired data has therefore been split into 10 data sets, and the data analysis performed separately for each of these data sets before being combined to final results. The total beam time over the 6 months period was approximately 300 hours. For further information see~\cite{laursen16,laursenPhd}.

%% file: analysis.tex
\section{Data analysis and results}
\label{sec:analysis}

The data analysis is structured in the following way. First events where three $\alpha$-particles are detected in coincidence are identified. By analysing the total centre of mass energies of these events a spectrum of final states populated by $\gamma$-decay of the 16.11~MeV state is generated. Then the partial decay widths and reduced transition strengths are deduced for the observed transitions based on eq.~\ref{gammagamma1}.

\subsubsection*{Identification of 3$\alpha$ events}
True 3$\alpha$ events are identified through a chain of steps. The main goal of these steps is to remove random coincidence events where one or more $\alpha$ particles are recorded in coincidence with either a noisy signal, an elastically scattered proton, or an $\alpha$ particle from a different breakup process. 

First, the TDC signals are used to reduce the number of random coincidences. Next, the front and back energies recorded in a detector are required to match within 150\,keV. The front-back matching procedure reconstructs summing and sharing events, the latter resulting from interstrip hits where the electron-hole pairs created by one particle are collected in two neighbouring strips. The summing contribution is significant in the $^8$Be(gs) channel due to the small opening angle between the secondary $\alpha$ particles. We then use that the total center-of-mass momentum is zero for a genuine triple-$\alpha$ breakup, and apply cuts on the sum of the $\alpha$ particle momenta as well as on the individual momentum components. These cuts can be used to pick out the $\gamma$-delayed events because the $\gamma$-decay induces a negligible recoil to the system. We also require the angles between the three $\alpha$ particles to sum to $360\degree$, and that the breakup must occur in one plane. For both these cuts a margin of $10\degree$ is allowed. The effect of the cuts is such that the TDC cut reduces the number of triple events by a factor 100 at the lowest energies and a factor 5-10 at 8-10 MeV sum energy, while the momentum and sum angle cuts remove an additional approximately 100 events.  Further information on the analysis procedure can be found in~\cite{alcorta09,laursen16,laursenPhd}. 

The events left over from imposing these analysis steps can be assumed to be true 3$\alpha$ events. It is useful to divide these events into two groups defined by the breakup kinematics. By calculating the relative energy for each pair of $\alpha$-particles, decays proceeding via the ground state of $^8$Be can be labelled as $^8$Be(gs). The remaining events are labelled as $^8$Be(exc). This step is discussed in greater detail in~\cite{laursen16}. 

\begin{figure}
\centering
\includegraphics[width=0.5\textwidth]{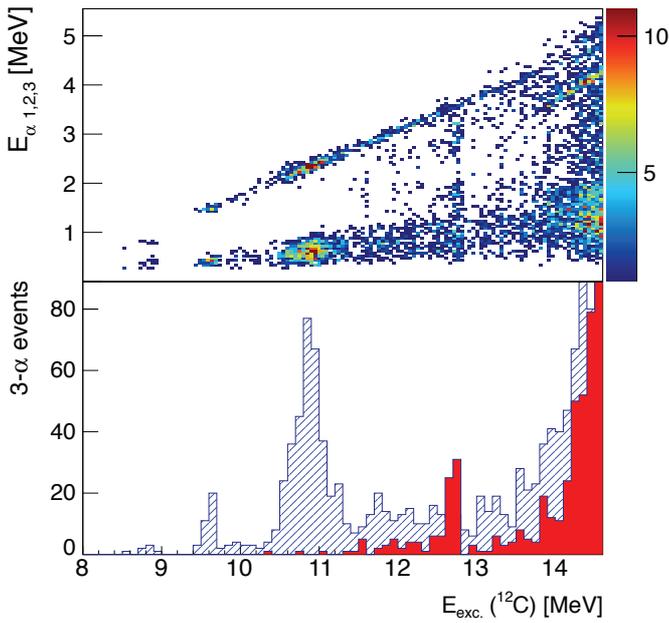}
\caption{\label{fig:gamma}(Color online) \textit{Bottom panel}: $^{12}$C energy spectrum for all data with the excitation energy along the abscissa. The hatched (blue) histogram is gated on the $^{8}$Be ground state and the filled (red) histogram represents decays proceeding through the first excited state in $^8$Be. \textit{Top panel}: two dimensional plot with the individual $\alpha$ particle energies, $E_{\alpha,i}$, along the ordinate and the excitation energy along the abscissa.}
\end{figure}

The $^{12}$C excitation energy spectrum, for all acquired data, is shown in the lower panel of Fig. \ref{fig:gamma}. The spectrum consists of three peak and an unresolved contribution. The peaks can be identified as the known levels $3^-$ ($E = 9.64\,\textrm{MeV, }\Gamma = 48(2)\,\textrm{keV}$ \cite{alcorta12,kokalova13}), $1^-$ ($E = 10.85\,\textrm{MeV, }\Gamma = 272(5)\,\textrm{keV}$ \cite{alcorta12}) and $1^+$ ($E = 12.71\,\textrm{MeV, }\Gamma = 18(5)\,\textrm{eV}$ \cite{ajzenberg90}). The transition to the $1^-$ state, and the unresolved contribution, are observed here for the first time. The fact that the transition to the $1^-$ state was missed in the previous studies using conventional $\gamma$-spectroscopy illustrates the advantages of the method of indirect $\gamma$ detection for identifying transitions to broad final states.

The excitation energy spectrum in Fig. \ref{fig:gamma} is divided into two types; one for the $^8$Be(gs) channel (hatched), and a second for the $^8$Be(exc) channel (filled). Thus, the hatched histogram illustrates transitions to states of natural parity, in agreement with our spin-parity assignment for the peaks at 9.6\,MeV and 10.8\,MeV. Natural parity events can also be identified around 10\,MeV and between 11.5-13.0\,MeV. These events cannot be linked to any known resonances in $^{12}$C. 

The two decay channels $^8$Be(gs) and $^8$Be(exc) can also be identified on the upper panel of Fig. \ref{fig:gamma} which shows a scatter plot of $^{12}$C excitation energies $E_\textrm{exc}$ determined from the total 3$\alpha$ decay energy versus the individual $\alpha$-particle energies $E_{\alpha,i}$. The $^8$Be(gs) channel can be identified from the two-body kinematics relation
\begin{align}
E_{\alpha,1} = \tfrac{2}{3}\left( E_\textrm{exc.} - E_{^8\textrm{Be(gs)}} \right),
\end{align}
which describes a linear relation between $E_{\alpha,i}$ and $E_\textrm{exc}$ corresponding to the diagonal in the top panel of Fig. \ref{fig:gamma} extending from the lower left to the upper right. 

Transitions to states that decay through the $^8$Be(exc) channel are also identified in Fig. \ref{fig:gamma}. The 12.71\,MeV ($1^+$) state, which only decays in the $^8$Be(exc) channel due to angular momentum and parity conservation, stands out most significantly, while events are also observed at 11.5-12.5\,MeV. These events could result from the known $2^-$ state at 11.83\,MeV \cite{ajzenberg90}, but may also represent a $^8$Be(exc) branch from the same source that gives rise to the $^8$Be(gs) events in these energy bins.  

\subsubsection*{Extraction of partial widths}
\label{subsec:partialwidths}
The partial decay widths can be calculated from eq.~\ref{gammagamma1}. For the total width we use the value discussed in section~\ref{sec:2p-review}, and the branching ratios we determine as:
\small{\begin{align}
B_{\gamma,i} = \frac{N_{\gamma,i}^\textrm{tot}}{N_{16.11}^\textrm{tot}} = \frac{ N_{\gamma,i}^\textrm{obs} / \epsilon_{\gamma,i}}{ N_{16.11,\textrm{gs}}^{\textrm{obs},3\alpha} / \epsilon_\textrm{gs} + N_{16.11,\textrm{exc}}^{\textrm{obs},3\alpha} / \epsilon_\textrm{exc} },
\label{eq:Gammagamma}   
\end{align}}\normalsize
where  $N_{\gamma,i}^\textrm{tot}$ is the total number of $\gamma$ decays in the $i$'th transition and $N_{16.11}^\textrm{tot}$ is the number of decays in total. $N_{\gamma,i}^\textrm{obs}$ is the observed number of $i$ transitions, $N_{16.11,\textrm{gs}}^{\textrm{obs},3\alpha}$ is the observed number of decays in the $^8$Be(gs) channel, $N_{16.11,\textrm{exc}}^{\textrm{obs},3\alpha}$ is the observed number of decays in the $^8$Be(exc) channel, and $\epsilon_{\gamma,i}$, $\epsilon_\textrm{gs}$ and $\epsilon_\textrm{exc}$ are the corresponding detection efficiencies. 
The detection efficiencies are strongly dependent on the 3$\alpha$-breakup mechanism, and also depend on the detailed analysis cuts placed on the detected events. Hence, the efficiencies must be determined by Monte-Carlo simulations.

Due to changes in the setup, such as in the detection geometry and the gradual adsorption on the target foil~\cite{laursen16}
, the efficiencies vary and are calculated for each of the 10 data sets separately. The calculations are based on $3\times10^6$ simulated breakup events and the efficiencies for detecting all three $\alpha$ particles range between 11-15\% and 0.2-0.8\% for the $^8$Be(gs) and $^8$Be(exc) decay channels, respectively. For $\gamma$ decays the appropriate phase-space factor, $E_\gamma^{2L+1}$, replaces the entrance penetrability in the Breit-Wigner function describing the daughter level. Next, experimental effects are taken into account. These include the geometry of the detection system and the response of the individual detectors. Finally, the simulated data is passed to the same analysis routines as applied to measured data hence accounting for any bias introduced by cuts and gates applied in the analysis procedure. 
\begin{table*}[ht]
\centering
\caption{\label{tab:sim_gamma_def}The five groups that efficiency simulation are performed for. Group 4 is divided into 6 intervals, each 250\,keV wide. The simulations in groups 4 and 5 are divided into two sub-groups assuming M1 and E1 decays to $2^+$ and $3^-$ states, respectively. Individual simulations are performed for all 10 data sets. $l_1$ and $l_2$ are the orbital angular momenta of the primary and secondary $\alpha$-decay of the $^{12}$C states.}
\begin{tabular}{cccccc} 
\toprule
Group & Decay group & $^8$Be breakup ch. [gs/exc] & E$L$/M$L$ & $l_1$ & $l_2$ \\ \hline
1 & $1^+$                          &    exc      &   M1         &  2  &  2 \\
2 & $1^-$                          &    gs       &   E1          &  1  &  0 \\
3 & $3^-$                          &    gs       &   E1          &  3  &  0 \\
4 & 11.5-13.0\,MeV ($2^+,3^-$)     &    gs       &   M1,E1   &  2,3  &  0 \\
5 & 9.8-10.25\,MeV ($2^+,3^-$)     &    gs       &   M1,E1   &  2,3  &  0 \\	
\bottomrule
\end{tabular}
\end{table*}

Uncertainties on trigger thresholds and deadlayer and target foil thicknesses affect the low energy detection efficiency strongly. This is particularly important for $\gamma$-delayed triple-$\alpha$ emissions where $\alpha$ particles of very low energy are emitted. In addition, geometry uncertainties strongly influence the efficiency for detecting decays in the $^8$Be(exc) channel. This is due to the large breakup angle between the secondary $\alpha$ particles, making it difficult to observe both of them with two detectors. These effects are varied within their estimated uncertainties and are responsible for the systematic errors on the final results. Compared to this, the systematic uncertainty coming from the description of the direct $\alpha$-decay of the 16.11 MeV state \cite{laursen16}, which is used to normalise the branching ratios of the $\gamma$-transitions, is negligible.

\subsubsection*{Reduced transition strengths}
The observed partial transition widths are calculated according to Eqs. (\ref{gammagamma1}) and (\ref{eq:Gammagamma}) and also converted to reduced transition strengths. This is done both for the identified transitions to the $3^-$, $1^-$ and $1^+$ levels and to the unidentified natural parity groups at 9.8-10.3\,MeV and 11.5-13.0\,MeV. The latter group is divided into six sub-groups, each 250\,keV wide. Table \ref{tab:sim_gamma_def} defines the groups in terms of physics input to the efficiency and transition strength calculations. Groups 4 and 5 are further divided into two sub-groups, one assuming $2^+$ as daughter level spin-parity and a second assuming it to be $3^-$.
These choices are motivated by an assumption that M1 and E1 are the dominating multipolarities, and that no more 1$^-$ states are expected in this energy region. The assumption of the J$^{\pi}$ of the populated state affects slightly the estimated detection efficiencies due to the difference in the shape of the phasespace factor for different multipolarities. Therefore the calculated partial widths are slightly different for the two cases.   

In the case of the $1^+$ level, which only decays through the $^8$Be(exc) resonance, the breakup is treated as described in \cite{fynbo03}, which is identical to model 4 in~\cite{laursen16}. For the remaining groups the breakup is assumed to proceed via the $^8$Be(gs), which is the only known decay route for the $3^-$ and $1^-$ levels (see also \cite{alcorta12}). 

Our recommended transition widths and reduced strengths for the $1^+$, $1^-$ and $3^-$ states are shown in Table \ref{tab:res_gamma_1} while Table \ref{tab:res_gamma_2} shows the corresponding results for the broad natural parity distributions in groups 4 and 5. Due to the very low number of observed $\gamma$ decays in each of the ten data sets (typically below 10 decays for the $3^-$ and $1^+$ levels, 40 decays for the $1^-$ level and even lower for decays groups 4 and 5) a maximum likelihood analysis is performed in order to calculate the average values. Tables \ref{tab:res_gamma_1} and \ref{tab:res_gamma_2} also list the observed $\gamma$ widths in units of the Weisskopf single-particle estimates. Compared to the Weisskopf estimates our widths are generally small with E1 transitions at the per mille level and the M1 transitions at the 10\% level. 

\begin{table*}
\centering
\caption{\label{tab:res_gamma_1} Measured $\gamma$-widths averaged over all data sets (see text) for transitions to the $1^+$, $1^-$ and $3^-$ levels and corresponding reduced matrix elements. The latter are calculated assuming the lowest possible multipolarity. The second column gives the total number of observed $\gamma$ decays. Column 4 shows the width values in units of the corresponding Weisskopf estimates. The statistical (first bracket) and systematic (second bracket) errors are indicated separately. }
\begin{tabular}{cccccc} 
\toprule
 & N$_\textrm{Tot}$ & $\Gamma_\gamma$ [eV] &  [W.u. $\times 10^{-3}$]  & $B\left(\textrm{E}1\right)$ [$e^2\,\textrm{fm}^2\times10^{-3}$]   & $B\left(\textrm{M}1\right)$ [$\mu_\textrm{N}^2$]   \\  \hline
$1^+$  & 71  & 0.14(2)(5)  & 170 & - & 0.31(4)(11)  \\
$1^-$  & 365 & 0.48(4)(11) & 9.3 & $3.2(3)(7)$ & - \\
$3^-$  & 36  & 0.33(8)(8)  & 3.4 & $1.2(3)(3)$ & - \\
\bottomrule
\end{tabular}
\end{table*}

For the $^8$Be(exc) decay events at 11.5-12.5\,MeV there can be two different interpretations. They may result from the same state(s) that gives rise to the natural parity distribution in this region. Assuming this, the $^8$Be(exc) branching ratio for the natural parity structure is calculated to $47\%$. This is a large value compared to the actual numbers observed in the ground and excited state channels, which are 131 and 32, respectively, but may be explained by the much lower detection efficiency for $^8$Be(exc) decays. Alternatively, they may result from unnatural parity states in this energy region. One candidate for this could be the 11.83 (2$^-$) resonance, which could be populated by an E1 transition.

\begin{table*}[ht]
\centering
\caption{\label{tab:res_gamma_2}Measured $\gamma$ widths averaged over all data sets (see text) for transitions to the natural parity distributions at 10.0\,MeV and at 11.5-13.0\,MeV. The second column gives the total number of observed $\gamma$ decays. Columns 3 and 6 give the $\gamma$ widths for the $2^+$ and $3^-$ cases, respectively, while columns 4 and 7 list the width values in units of the corresponding Weisskopf estimates. Columns 5 and 8 list the reduced matrix elements. The statistical (first bracket) and systematic (second bracket) errors are indicated separately.}
\begin{tabular}{ccccccccc} 
\toprule
&&\multicolumn{3}{c}{$2^+$}&&\multicolumn{3}{c}{$3^-$} \\ \cline{3-5}\cline{7-9}
$\Delta$ E [MeV] & N$_\textrm{Tot}$ & \multicolumn{2}{c}{$\Gamma_\gamma$} & $B\left(\textrm{M}1\right)$ && \multicolumn{2}{c}{$\Gamma_\gamma$} & $B\left(\textrm{E}1\right)$ \vspace{0.05cm}\\ \cline{3-4}\cline{7-8}
&& [eV] & [W.u. $\times 10^{-3}$]  & [$\mu_\textrm{N}^2$] && [eV] & [W.u. $\times 10^{-3}$] & [$e^2 (\textrm{fm})^2\times10^{-3}$]  \\ \hline
9.8-10.3           &    15     &   0.018(7)(6)  & 0.22 & 0.007(3)(2) && 0.018(8)(6)  & 3.8& 0.08(3)(3) \\
11.5-11.75      &    32     &   0.021(5)(5)  &  0.66 & 0.020(5)(5) && 0.021(5)(5)  & 11.2& 0.22(5)(5) \\
11.75-12.00    &    35     &   0.019(4)(5)  & 0.70  & 0.022(5)(6) &&  0.019(4)(5) & 12.0& 0.24(5)(6) \\
12.00-12.25    &    33     &   0.015(4)(4)  &  0.67 & 0.021(6)(6) && 0.015(4)(3)  & 11.4& 0.23(6)(4) \\
12.25-12.50    &    31     &   0.012(3)(3)  &  0.65 & 0.020(5)(5) &  & 0.014(3)(3) & 13.0& 0.26(6)(6) \\
12.50-12.75    &    31     &   0.011(3)(3)  &  0.73 & 0.023(6)(6) && 0.012(3)(2)  & 13.7& 0.27(7)(5) \\
12.75-13.00    &    21     &   0.007(3)(2)  &  0.58 & 0.018(3)(2) && 0.008(3)(2)  & 11.4& 0.23(9)(6) \\
\bottomrule
\end{tabular}
\end{table*}

%% file: discussion.tex
\section{Discussion and Outlook}
\label{sec:discussion}
Use of indirect detection of $\gamma$-decays has allowed identification of considerable more strength then found earlier. 

The partial widths for the previously observed transitions to the 9.64\,MeV (3$^-$) and 12.71\,MeV (1$^+$) states are consistent with the previous estimates given in table~\ref{tab:partialwidths-review}. A stronger transition to the known 10.84\,MeV (1$^-$) state was identified for the first time. 

The statistics in the unresolved part of the spectrum is too little to make any firm conclusions. However, an obvious interpretation for the observed strength at 10\,MeV is the $2^+$ structure recently observed in this region \cite{freer09,zimmerman11,itoh11,smit12,zimmerman13}, while the strength at 11.5-13.0\,MeV is close to the $2_2^+$ position found in $\beta$-decay experiments \cite{hyldegaard10}, but this region also overlaps with the position of the $3^-$ state suggested by \cite{freer07}. The strength in the $^8$Be(exc) channel in the 11.5-12.5~MeV region is most likely due to feeding to the known 2$^-$ state at 11.83~MeV~\cite{ajzenberg90} because it is unlikely that a natural parity resonance in that energy region would have such a large partial decay branch to the $^8$Be(exc) channel.

An alternative interpretation of the unresolved strength is direct capture to the $^8$Be+$\alpha$ continuum. However, Kelley {\it et al.} \cite{kelley00} have estimated direct capture to be at least two orders of magnitude less than resonant capture for the 16.11\,MeV 2$^+$ state. Also, a contribution from $\gamma$-decay of the $^8$Be 2$^+$ resonance to the ground state of $^8$Be could contribute to the unresolved strength,
\begin{equation*}
    ^{12}\textrm{C(16.11MeV)} \rightarrow {^8}\textrm{Be(2$^+$)}+\alpha \rightarrow {^8}\textrm{Be(0$^+$)}+\gamma + \alpha \,.
\end{equation*}
However, the $\gamma$-branching ratio of the $^8$Be 2$^+$ resonance is estimated to be less than 10$^{-8}$~\cite{wiringa00,datar05}, and therefore less than one event of this type is expected to contribute to the data.

The measured partial widths to the resolved and unresolved resonances should provide strong constraints on theoretical models predicting resonances in this important region of the spectrum of $^{12}$C. Such calculations are strongly encouraged.

The experiment presented here applied a relatively simple detection setup of only two DSSSDs. The use of a larger setup covering a larger solid angle would increase the detection efficiency for 3$\alpha$-events considerably and therefore provide more statistics. This would shed more light on the unresolved parts of the spectrum observed here.